\documentclass[sigconf]{acmart}
\pdfoutput=1
\usepackage{multirow}
\usepackage{bbding}
\usepackage{pifont}
\usepackage{wasysym}

\usepackage{amssymb}
\usepackage{algorithm}
\usepackage[noend]{algpseudocode}

\AtBeginDocument{%
  \providecommand\BibTeX{{%
    \normalfont B\kern-0.5em{\scshape i\kern-0.25em b}\kern-0.8em\TeX}}}

\setcopyright{acmcopyright}
\copyrightyear{2022}
\acmYear{2022}
\acmDOI{XXXXXXX.XXXXXXX}
\acmConference[MM '22]{Proceedings of the 30th ACM International Conference on Multimedia}{October 10--14, 2022}{Lisbon, Portugal}

\begin{document}

\title{HERO: HiErarchical spatio-tempoRal reasOning with Contrastive \\ Action Correspondence for End-to-End Video Object Grounding}

\author{Mengze Li}
\affiliation{%
  \institution{Zhejiang University}
  \city{Hangzhou}
  \state{Zhejiang}
  \country{China}}
\email{mengzeli@zju.edu.cn}

\author{Tianbao Wang}
\affiliation{%
  \institution{Zhejiang University}
  \city{Hangzhou}
  \state{Zhejiang}
  \country{China}}
  
\author{Haoyu Zhang}
\affiliation{%
  \institution{Zhejiang University}
  \city{Hangzhou}
  \state{Zhejiang}
  \country{China}}

\author{Shengyu Zhang}
\affiliation{%
  \institution{Zhejiang University}
  \city{Hangzhou}
  \state{Zhejiang}
  \country{China}}

\author{Zhou Zhao}
\authornote{Corresponding authors.}
\affiliation{%
  \institution{Zhejiang University}
  \institution{Shanghai Institute for Advanced Study of Zhejiang University}
  \city{Hangzhou}
  \state{Zhejiang}
  \country{China}}
\email{zhaozhou@zju.edu.cn}

\author{Wenqiao Zhang}
\authornotemark[1]
\affiliation{%
  \institution{National University of Singapore}
%   \city{Singapore}
%   \state{Zhejiang}
  \country{Singapore}}
\email{wenqiaozhang@zju.edu.cn}

\author{Jiaxu Miao}
\authornotemark[1]
\affiliation{%
  \institution{Zhejiang University}
  \city{Hangzhou}
  \state{Zhejiang}
  \country{China}}
\email{jiaxu.miao@yahoo.com}

\author{Shiliang Pu}
\affiliation{%
  \institution{Hikvision}
  \city{Hangzhou}
  \state{Zhejiang}
  \country{China}}

\author{Fei Wu}
\authornotemark[1]
\affiliation{%
  \institution{Shanghai Institute for Advanced Study of Zhejiang University}
  \institution{Shanghai AI Laboratory}
%   \institution{Institute of Artificial Intelligence, Zhejiang University}
%   \city{Hangzhou}
  \state{Shanghai}
  \country{China}
  }
\email{wufei@zju.edu.cn}

\renewcommand{\shortauthors}{Mengze Li, et al.}
\newcommand{\ie}{\textit{i}.\textit{e}.}

\begin{abstract}
\underline{V}ideo \underline{O}bject \underline{G}rounding (\emph{VOG}) is the problem of associating spatial object regions in the video to a descriptive natural language query. This is a challenging vision-language task that necessitates constructing the correct cross-modal correspondence and modeling the appropriate spatio-temporal context of the query video and caption, thereby localizing the specific objects accurately. In this paper, we tackle this task by a novel framework called \underline{\textbf{H}}i\underline{\textbf{E}}rarchical spatio-tempo\underline{\textbf{R}}al reas\underline{\textbf{O}}ning (\textbf{HERO}) with contrastive action correspondence. We study the VOG task at two aspects that prior works overlooked: (1) \emph{Contrastive Action Correspondence-aware Retrieval.} 
Notice that the fine-grained video semantics (\emph{e.g.}, multiple actions) is not totally aligned with the annotated language query (\emph{e.g.}, single action), we first introduce the weakly-supervised contrastive learning that classifies the video as action-consistent and action-independent frames relying on the video-caption action semantic correspondence. Such a design can build the fine-grained cross-modal correspondence for more accurate subsequent VOG. (2) \emph{Hierarchical Spatio-temporal Modeling Improvement}. While transformer-based VOG models present their potential in sequential modality (\emph{i.e.}, video and caption) modeling, existing evidence also indicates that the transformer suffers from the issue of the insensitive spatio-temporal locality. Motivated by that, we carefully design the hierarchical reasoning layers to decouple fully connected multi-head attention and remove the redundant interfering correlations. Furthermore, our proposed pyramid and shifted alignment mechanisms are effective to improve the cross-modal information utilization of neighborhood spatial regions and temporal frames.
We conducted extensive experiments to show our HERO outperforms existing techniques by achieving significant improvement on two benchmark datasets.
\end{abstract}

\begin{CCSXML}
<ccs2012>
 <concept>
       <concept_id>10010147.10010178.10010179</concept_id>
       <concept_desc>Computing methodologies~Natural language processing</concept_desc>
       <concept_significance>500</concept_significance>
       </concept>
   <concept>
       <concept_id>10010147.10010178.10010224</concept_id>
       <concept_desc>Computing methodologies~Computer vision</concept_desc>
       <concept_significance>500</concept_significance>
       </concept>
\end{CCSXML}

\ccsdesc[500]{Computing methodologies~Natural language processing}
\ccsdesc[500]{Computing methodologies~Computer vision}

\keywords{multi-modal video object grounding, weakly-supervision, multi-head self attention}

\maketitle

\section{Introduction}
Visual object grounding aims to automatically detect the objects described by the natural language text from visual information, which involves mutual understanding and reasoning across different modalities. It is a prominent and fundamental vision-and-language task that has received increasing attention recently \cite{chen2018temporally, anne2017localizing, mao2016generation, ye2021one, tang2021human}. Generally, visual object grounding has two directions that are \underline{I}mage \underline{O}bject \underline{G}rounding (\emph{IOG}) \cite{anne2017localizing, mao2016generation, ye2021one} and \underline{V}ideo \underline{O}bject \underline{G}rounding (\emph{VOG}) \cite{tang2021human, zhang2020object, su2021stvgbert}. In the past few years, benefitted from the rapid developments of deep learning \cite{wu2020biased, ZhangYYLFZC022, gan2021dependency}, the prevailing \emph{IOG} methods prefer to ground the object words by constructing the implicitly cross-modal correspondence between image and query text, which has made remarkable progress~ \cite{yu2018mattnet, ye2021one}. 

\begin{figure}[t]
  \centering
  \includegraphics[width=\linewidth]{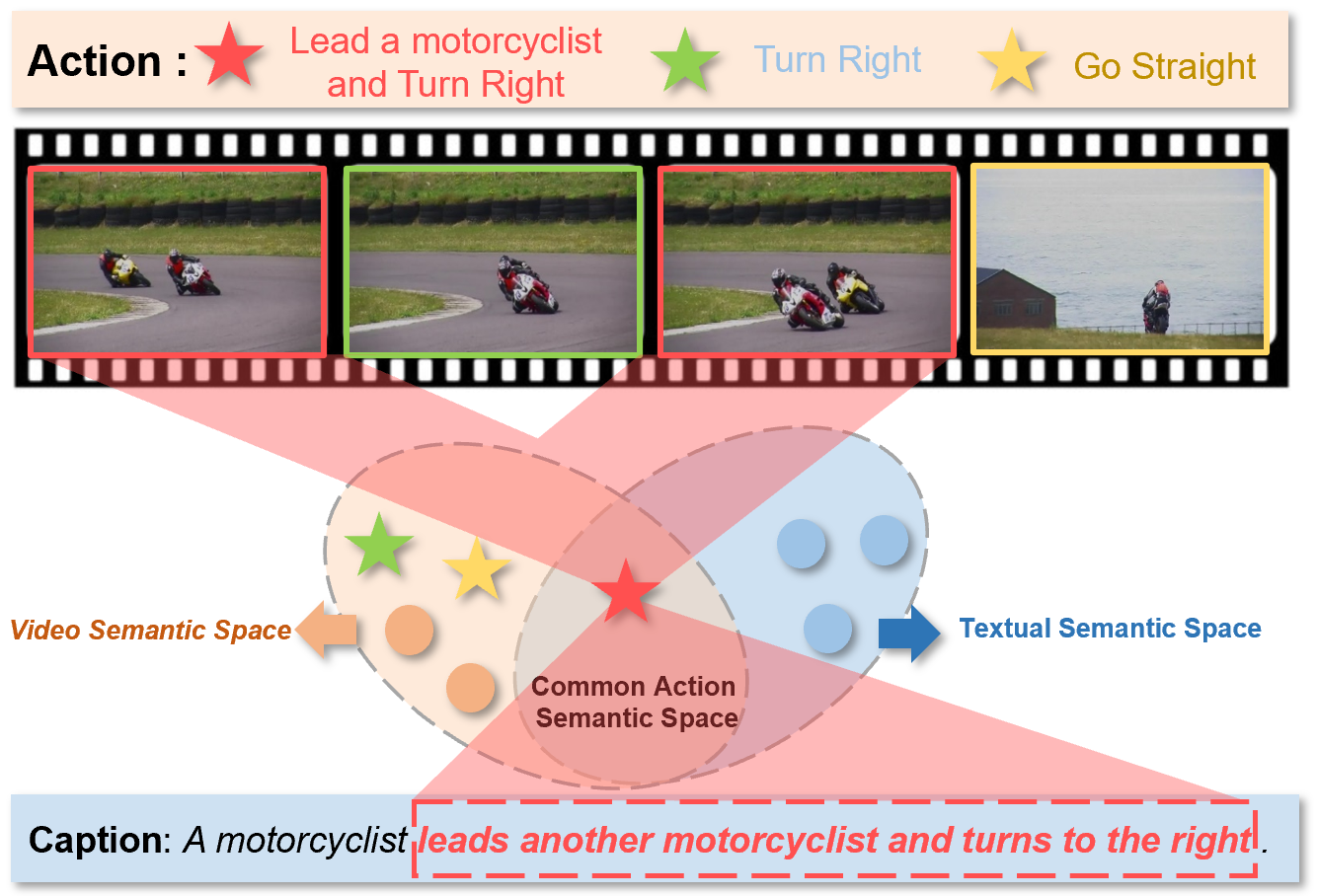}
  \caption{An example of the incomplete cross-modal semantic correspondence issue in video object grounding task. Pentagrams denote the action in video and caption semantic space. Circles represent the other semantics, \ie, objects.}
  \label{figure1-1}
\end{figure}

Compared to IOG, VOG is more practical but more challenging. As shown in Fig.~\ref{figure1-1}, VOG requires comprehending the complicated semantics from video redundantly sequential video frames and further finding the appropriate object regions according to the concisely descriptive caption.
Broadly, most of the contemporary VOG approaches prefer to extend the IOG-based cross-modal correspondence learning paradigm,  i.e., processing the sequential information by
strong representational learning models (\emph{e.g.}, RNN~ \cite{zaremba2014recurrent, hochreiter1997long} or transformer~ \cite{vaswani2017attention}) and then learn the implicitly corresponding evidence of cross-modal correlations for the precise object grounding. Despite intuition, several critical factors for VOG have been ignored by existing methods, which can be summarized as two aspects: 
(A)~\textbf{Incomplete Cross-modal Semantic Correspondence.} Compared with the image \cite{gan2020deep, ge2021structured, zhang2020joint}, the video often contains more spatio-temporal semantic information (\emph{e.g.}, diverse actions) than the textual query (\emph{e.g.}, single action), some of the them even seems irrelevant or contrary to the action text intuitively. For instance, as shown in Figure~\ref{figure1-1}, the motorcyclist \emph{turns right with another motorcyclist}, and then \emph{goes straight}, not totally matched with the language description of \emph{turning right and leading a motorcyclist}. Put it another way, the fine-grained semantics in the video are not always consistent with the coarse-grained textual query. Using the simple IOG extension is hard to satisfy the complex VOG task, which may 
result in an incomplete semantic cross-correlation and lead to a decrease in model visual detection ability. There is a straightforward way that we can introduce more human annotations to alleviate these incomplete issues \cite{li2017tracking, yamaguchi2017spatio}.
However, it comes with labor-intensive labeling and time-consuming model training. (B)~\textbf{Insensitive Spatio-Temporal Locality Inference.} While the transformer has shown superior modeling ability for visual-text sequential learning tasks such as the video question answer and captioning \cite{zhang2021image, ji2021improving, cornia2020meshed, zhou2021trar}, there are a few attempts~ \cite{su2021stvgbert, li2022end} using the transformer to tackle the VOG task. However, recent quantitative analyses~ \cite{xu2021vitae} indicate the pure vision transformers are lacking convolutional inductive biases (\emph{e.g.}, translation equivariance \cite{worrall2017harmonic}), which may lead the insensitive spatial locality in cross-modal modeling. Moreover, this situation is more serious and undesirable in video modeling, i.e., besides the spatial aspect, the temporal semantic modeling by the transformer also meet the insensitive locality issue. As a result, the transformer's reasoning abilities for the VOG task might be limited if it misses this spatio-temporal local sensitivity.
Based on this insight, our goal is to break away from VOG conventions and conduct an end-to-end fine-grained cross-modal correspondence and hierarchical interaction reasoning to complete the VOG task better.

To address these challenges, we devise an \underline{\textbf{H}}i\underline{\textbf{E}}rarchical spatio-tempo\underline{\textbf{R}}al reas\underline{\textbf{O}}ning model (\textbf{HERO}) with contrastive action correspondence for end-to-end VOG task. The model includes the following targeted designs: 
(A)~\textbf{Cross-modal Semantic Alignment}. According to the input language query, the HERO learns to divide the video into two parts, textual action-consistent frames and action-independent frames, by the weakly-supervised method. Technically, contrastive learning is adopted from global, clip, and frame levels to train the cross-modal action semantic matching ability of the model. 
(B)~\textbf{Hierarchical Spatio-Temporal Locality Enhancement}.
We propose the hierarchical reasoning network to simplify redundant attention among adjacent frame regions, which interferes with the local sensitivity of the traditional transformer layer. Furthermore, the pyramid and shifted locality alignment mechanisms are introduced to this module. Such novel blocks explicitly highlight the local correlation of the 
regions among adjacent or nearby frames, to effectively enhance the attribute (\emph{e.g.}, object) visual spatial- and action visual temporal- local sensitivity.

We conduct experiments on two widely used benchmark datasets. The results demonstrate the effectiveness of our HERO over the state-of-the-art. Extensive experiments for the video object grounding task including ablation studies and case studies demonstrate the significant merits of the HERO model. Our contributions are summarized as follows:

\begin{itemize}
  
  \item We propose a novel pre-processing learning mechanism called contrastive action correspondence-aware retrieval that is able to construct more fine-grained cross-modal correspondence through multi-granular contrastive learning for accurate video object grounding.

  \item We tackle the critically insensitive spatio-temporal locality of transformer in video object grounding via two aspects: (1) developing a hierarchical reasoning network to remove the redundant correlations among adjacent frame regions; (2) introducing the novel pyramid and shifted locality alignment block to utilize the attribute spatial locality and the action temporal locality better.

  \item The consistent superiority of the proposed HERO is demonstrated on the two benchmark datasets that outperforms the existing SOTA methods by a large margin.

\end{itemize}

\section{Related Work}

{\bfseries Visual Object Grounding.} Visual object grounding is to localize the target objects in the image described by the language query. In recent years, deep learning has been proved to be effective \cite{anpeng2022instrumental, guo2022collaborative, ZhangTZYKJZYW20, guo2021semi, zhang2019frame}. Thus, some researchers introduce it into the visual task \cite{li2020ib, zhang2022boostmis, kong2022attribute, zhang2020relational, li2019walking}, such as the visual object grounding. Most methods follow the two-stage pipeline. Some early works \cite{mao2016generation, nagaraja2016modeling} use CNN to extract visual features from images and then use RNN for learning a probability of a referring expression. The later work \cite{nagaraja2016modeling} tries to measure the similarity between the objects and languages in a common feature space. \cite{nagaraja2016modeling} extracts subject appearance, location, and relationship of objects, and fuses them with the attention layer. Recently, researchers explore the way to represent the structure information of images as a graph to construct the relation reasoning process \cite{hu2019language, wang2019neighbourhood}. Different from \cite{hong2019learning}, which merely constructs the language-based graph to capture the semantics in expression, \cite{jing2020visual} uses joint reasoning over both the language graph and visual graph. 

One-stage approach \cite{ye2021one, yang2019fast} extracts the feature maps from the input image and views the object bounding box prediction as a pixel-level regression problem, different from the two-stage approaches generating a series of proposals explicitly. \cite{ye2021one} proposes a one-stage network that utilizes the scene graph of natural language as an extra input and incorporates a semantic reconstruction loss. \cite{liao2020real} proposes to transfer the problem of visual object grounding to a multi-model correlation filtering process and the authors solve it with a one-stage joint optimization paradigm.

{\bfseries Video Object Grounding.} Many researchers have been focusing their attention on video related tasks, among many multi-modal tasks \cite{zhang2022magic, wang2021weakly, li2020unsupervised,  zhang2021consensus, li2021adaptive, ZhangJWKZZYYW20}. 
From them, there are lots of existing works focusing on video temporal grounding \cite{han2021fine, hou2021conquer, li2022compositional, li2022dilated}. However, the video spatial grounding gets less attention, which localizes the target objects in the video given a language query. \cite{huang2018finding} explores the weakly-supervised video grounding framework. \cite{wang2021weakly} tries to ground the linguistic reference in the video with a MIL-based grounding approach. To achieve higher prediction accuracy, the researchers study video object grounding based on dense annotations. Video object detection \cite{lin2020dual, han2020exploiting} is closely related to video object grounding, but it localizes the object in a single-modal way. Several studies explore the video object tracking task with natural language sentences \cite{li2017tracking, ma2021capsule}. \cite{yang2020grounding} grounds the target region of a frame in the video according to the language description, and then, tracks the target objects according to the prediction for the previous frames. Finally, it utilizes the RT-integration method to judge whether the grounded bounding box accurately highlights the language referred objects and corrects grounding failures.

There are many two-stage approaches for the video object grounding task \cite{tang2021human, sadhu2020video, zhang2020does}. \cite{zhang2020does} proposes a two-stage framework that produces a series of object proposals by a pre-trained object detector in the first stage, and then, selects the target one from the candidates based on a spatio-temporal graph network in the second stage. In view of the superior performance of the transformer \cite{li2020multi, gan2020investigating}, some researchers try to use it to improve model performance.
\cite{tang2021human} uses the visual transformer to extract cross-modal representation for the human-center videos. \cite{sadhu2020video} applies self-attention to deal with the multi-modal features to encode the relation of the language and the visual objects.
Later researchers apply one-stage frameworks in the video object grounding task \cite{su2021stvgbert, li2022end}. \cite{su2021stvgbert} applies a transformer-based module to fuse the cross-modal information, which does not rely on the pre-trained object detector. \cite{li2022end} uses an information tree method to handle one-shot video object grounding tasks in an end-to-end manner, which could improve the representation ability under insufficient labeling.

\section{Method}
\subsection{Model Overview}

\subsubsection{Problem Formulation} Given natural language sentences $\textbf{S}=\{\textbf{S}_i\}_{i=1}^{N_S}$ and the set of videos $\textbf{V}=\{\textbf{V}_i\}_{i=1}^{N_V}$, the video object grounding task aims to localize the target objects $\textbf{O}=\{\textbf{O}_i\}_{i=1}^{N_O}$ in the videos. The $N_S$, $N_V$, and $N_O$ are the size of sentence set, video set and object set. It is needed to apply all the paired samples $((\textbf{V}, \textbf{S}), \textbf{O})$ to train our target grounding model, HERO. We define $\mathcal{M}$ as model with the initialized parameter $\Theta$. Then, the training process is to optimize the function:
\begin{equation}
  \mathcal{M}((\textbf{V}, \textbf{S}), \textbf{O}; \Theta) = \max\limits_\Theta \xi(\epsilon(\textbf{V}, \textbf{S}), \delta(\textbf{O})),
\end{equation}
where the functions $\epsilon(\textbf{V}, \textbf{S})$ and $\delta(\textbf{O})$
output the bounding boxes of the predicted and target objects, separately. The $\xi$ function calculates the coincidence of them.

\subsubsection{Pipeline of The Model} 
\begin{figure*}[h]
  \centering
  \includegraphics[width=\linewidth]{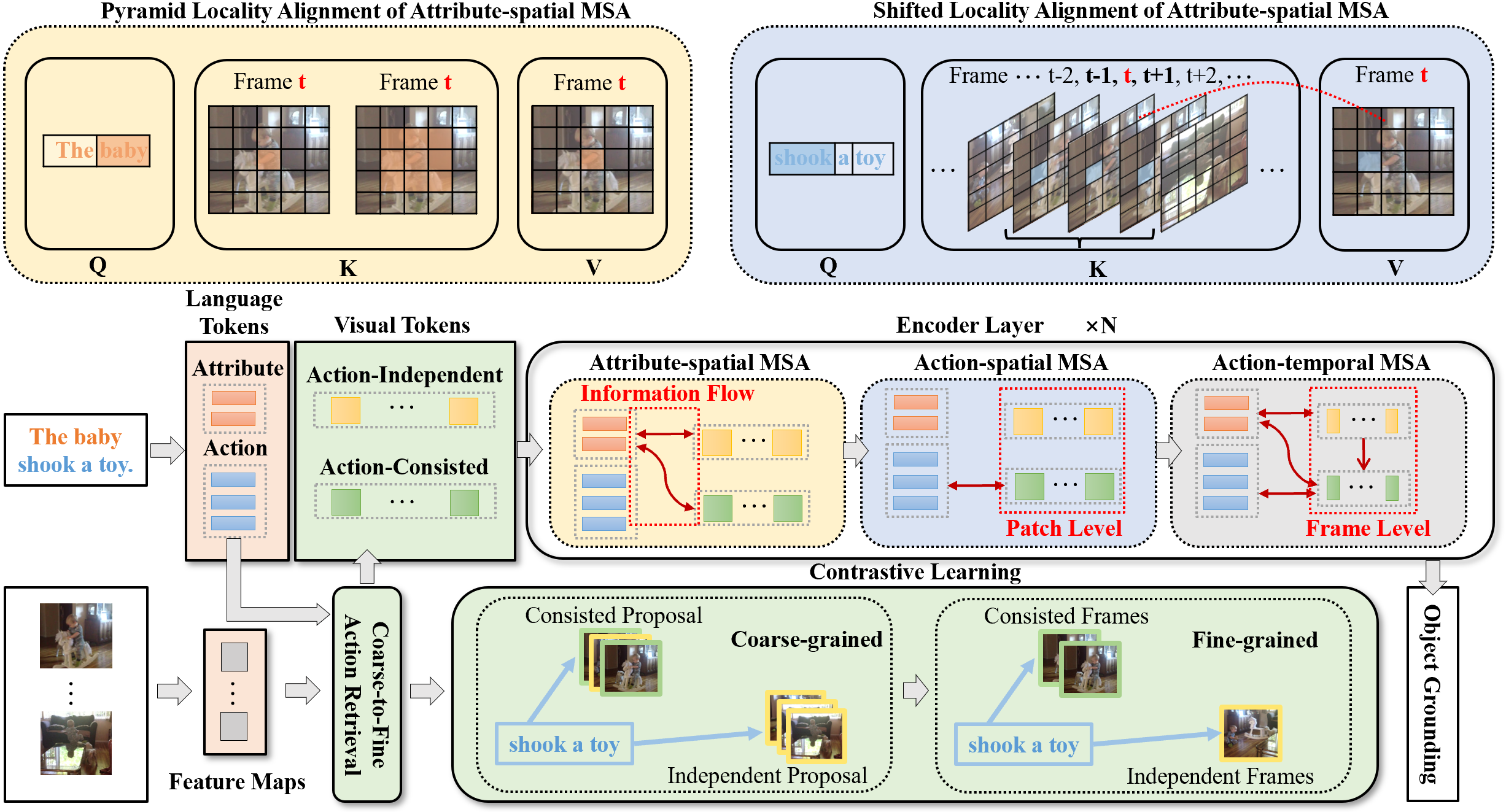}
  \caption{The overview  of our proposed \underline{\textbf{H}}i\underline{\textbf{E}}rarchical spatio-tempo\underline{\textbf{R}}al reas\underline{\textbf{O}}ning framework (HERO) with contrastive action correspondence for end-to-end video object grounding. Firstly, the Coarse-to-Fine Action Retrieval module distinguishes the video frames into action text consistent part and independent part. Then, the encoder layer reasons on the different video and text semantic parts step by step with three carefully designed \underline{M}ulti-head \underline{S}elf \underline{A}ttention (\emph{MSA}) layers. Notably, the pyramid and shifted alignment mechanisms are designed for Attribute-spatial MSA and Action-spatial MSA to fully use the spatial locality of the attribute semantics and the temporal locality of the action semantics.}
  \label{figure3-1}
\end{figure*}

As shown in Figure \ref{figure3-1}, after extracting features from video frames and input text with Resnet101 \cite{he2016deep} and RoBERTa \cite{liu2019roberta}, there are mainly two steps involved in the end-to-end modeling of HERO. The algorithm flowchart is shown in the Alg~\ref{alg3-1}.

\begin{algorithm}[t]
\renewcommand{\algorithmicrequire}{\textbf{Input:}}
\renewcommand{\algorithmicensure}{\textbf{Prepare}}
\caption{The inference process of the HERO model.}
\label{alg3-1}
% \textbf{Input: }private dataset $D$, public text dataset $T$, public image dataset $I$
\begin{algorithmic}[1]
\Ensure \\
initialize the \underline{C}oarse-to-\underline{F}ine \underline{A}ction \underline{R}etrieval module $(\textbf{CFAR})$ \\
initialize the \underline{AT}tribute-\underline{S}patial MSA $(\textbf{ATS})$, \underline{AC}tion-\underline{S}patial MSA $(\textbf{ACS})$, and \underline{AC}tion-\underline{T}emporal MSA $(\textbf{ACT})$ \\
initialize the \underline{D}ecoder \underline{L}ayer $(\textbf{DL})$ 
\end{algorithmic}

\begin{algorithmic}[1]
\Require a video $\textbf{V}$ and a natural language query $\textbf{S}$, which consists of the attribute description part $\textbf{S}_{at}$ and the action description part $\textbf{S}_{ac}$
\State $\textbf{V}_{at}, \textbf{V}_{ac} \gets CFAR(\textbf{V}, \textbf{S}_{at})$ 

\For {$i \gets 1 $ \textbf{to} $\textbf{N}_{en}$ }
\State $\textbf{V}, \textbf{S}_{at} \gets ATS(\textbf{V}, \textbf{S}_{at})$
\State $\textbf{V}_{ac}, \textbf{S}_{ac} \gets ACS(\textbf{V}_{ac}, \textbf{S}_{ac})$
\State $\textbf{V}, \textbf{S} \gets ACT(\textbf{V}, \textbf{S})$
\State $\textbf{V} \gets ff(\textbf{V})$, where $ff$ is the linear layer
\State $\textbf{S} \gets ff(\textbf{S})$
\EndFor

\For {$i \gets 1 $ \textbf{to} $\textbf{N}_{de}$ }
\State $\textbf{F} \gets DL(\textbf{V}, \textbf{S})$
\EndFor
\State $\textbf{P} \gets ff(\textbf{F})$, where $\textbf{P}$ is the predicted result

\end{algorithmic}
\end{algorithm}

\begin{itemize}
  \item Firstly, aiming at the mismatching of the action semantics between cross-modal information, we develop the cascaded action retrieval modules to distinguish the text action-consistent and action-independent video frames. The Coarse-grained Action Retrieval module finds out the most related video clip from lots of proposals. Furthermore, the Fine-grained Action Retrieval module will further choose the frames corresponding to the action text description from the selected video clip.
  \item Secondly, with the divided video frames and the corresponding language description, we fuse the multi-modal information step by step to avoid the interference of different semantics.
  Specifically, the key structure hierarchical \underline{M}ulti-head \underline{S}elf
  \underline{A}ttention (\emph{MSA}) modules consist of Attribute-spatial MSA, Action-spatial MSA and Action-temporal MSA. They reason on the multi-modal information from different perspectives.
  Notably, to fully use the spatial locality of attribute (\ie, object) semantics and the temporal locality of action semantics, we introduce the pyramid and shifted alignment structure to these MSA modules. 
\end{itemize}

\subsubsection{Prediction and Training} 
We follow the common prediction and training methods of visual transformer used in other object detection models \cite{kamath2021mdetr}. The transformer decoder outputs the possible prediction region features for each frame. For each possible region, a possibility and a bounding box are generated. We choose the bounding box with the highest possible value for each frame as the target box. 

During the training process, we calculate the possible prediction regions, firstly. The regions are matched with the target boxes, and the best match is chosen in each frame. Finally, use the match to train our HERO model. 

For the Coarse-to-Fine Action retrieval module, we apply the contrastive learning to train it from global, clip, and frame levels, together with the bounding box prediction training. 

\subsubsection{Formal Definition of Attention} Following the design of the previous attention structure \cite{vaswani2017attention}, we define the attention function as:
\begin{equation}
  SeAttn(\textbf{Q}, \textbf{K}, \textbf{V}; W) = W(\textbf{QK}^T)\textbf{V}.
\end{equation}
In it, $W$ is the trainable parameter. The $\mathbf{QK}^T$ measures how similarity between the elements. Based on the calculated similarity, the value of elements is fused to get a new one. 

\underline{M}ulti-head \underline{S}elf \underline{A}ttention (\emph{MSA}) is designed based on above structure. It projects the $\textbf{Q}, \textbf{K}, \textbf{V}$ onto different vectors. An attention function $SeAttn(\textbf{Q}, \textbf{K}, \textbf{V}; W)$ is applied for different projections. The output is transformed by a linear layer for the concatenation of all attention outputs:
\begin{equation}
  MSA(\textbf{Q}, \textbf{K}, \textbf{V}) = concat(\textbf{A}_1, ...,\textbf{A}_h)W^A, 
\end{equation}
\begin{equation}
  where \textbf{A}_i = SeAttn(\textbf{Q}W_i^Q, \textbf{K}W_i^K, \textbf{V}W_i^V; W_i).
\end{equation}

In it, the $W^A$, $W^Q$, $W^K$, and $W^V$ are learnable. 

\subsection{Contrastive Action Correspondence-aware Retrieval}
The input video often contains diverse actions. However, the corresponding text query is a coarse-grained description and often consists of a single action. Thus, they are not totally matched. To retrieve aligned cross-modal information,
in this section, we introduce the coarse-to-fine action retrieval modules, which distinguish the textual action relevant and irrelevant video semantics from the clip level and the frame level, respectively.

\subsubsection{Coarse-grained Action Retrieval}
Due to temporal semantic correlation, the textual action-consistent video frames are generally concentrated in a video clip. We devise the Coarse-grained Action Retrieval module to select it from the candidate proposals. Specifically, we represent the features of the frames output by the image encoder as $\textbf{F} = \{\textbf{f}^i\}_{i=1}^I$, and the candidate proposal features as $\textbf{P} = \{\textbf{p}^i\}_{i=1}^K$. They are input together to the Coarse-grained Action Retrieval module.

The structure of this module is similar to the transformer encoder, which consists of $2$ layers.
The multi-head attention layer is the key part, which is responsible for reasoning on the proposal features with the frame features. 
The previous multi-head attention layer analyzes all the input features together, indiscriminately. Due to different proposals corresponding to different video clips, such method is not suitable obviously. To solve the problem, we utilize the hard attention manner to fuse the candidate proposals with their corresponding video frames. Assume that candidate proposal $\textbf{p}^i$ is defined to represent the features of video clips from frame $m$ to frame $n$, then the fusion process is:

\begin{equation}
  \overline{\textbf{p}}^i = MSA(\textbf{p}^i, [\textbf{p}^i, \textbf{f}^m, \textbf{f}^{m+1}, ..., \textbf{f}^n], [\textbf{p}^i, \textbf{f}^m, \textbf{f}^{m+1}, ..., \textbf{f}^n]).
\end{equation}

Assume that the similarity between $\textbf{X}$ and $\textbf{Y}$ is defined as:

\begin{equation}
  Sim(\textbf{X}, \textbf{Y}) = \frac{\textbf{X}^T \textbf{Y}}{\|\textbf{X}\|_2 \|\textbf{Y}\|_2}.
\end{equation}
It is worth noting that $\textbf{X}$ and $\textbf{Y}$ are first projected to the same space with the multilayer perception before calculation

With the new candidate proposal features $\textbf{P}$ output by the Coarse-grained Action Retrieval module, we calculate the cosine similarity between them and the action text feature $\textbf{s}_{act}$:
\begin{equation}
  \theta^i = Sim(\textbf{p}^i, \textbf{s}_{act}),
\end{equation}
where the $MLP$ is the multilayer perception, the $\theta^i$ is the similarity between the proposal feature and the text feature.
The action text most relevant video proposal will be further analyzed in the next module. 

\subsubsection{Fine-grained Action Retrieval}
Note that not all the frames of the selected clip are consistent with the action language description. Thus, we design the Fine-grained Action Retrieval module to distinguish these frames.
With the proposal selected out by the Coarse-grained Action Retrieval module, we input its feature $\textbf{p}^c$ together with its corresponding frames' features $\textbf{F}_c = \{\textbf{f}^i\}_{i=m_c}^{n_c}$, and the textual action feature $\textbf{s}_{act}$. The proposal feature is updated with:
\begin{equation}
  \mathbf{\overline{p}}^c = MSA(MLP([\textbf{p}^c, \textbf{s}_{act}]), \textbf{F}_c, \textbf{F}_c),
\end{equation}
where the $MLP$ is the multilayer perceptron.

During the fusion process, the multi-head attention values $\textbf{V} = \{\textbf{V}_i\}_{i=1}^L$ are predicted for the frames, where $\textbf{V}_i = \{\textbf{v}_i^j\}_{j=1}^{n_c-m_c+1}$. We need to distinguish the text-consistent and independent frames according to these attention values.

Firstly, for each single-head attention, we normalize the values using the softmax function, and then, we sum them up and calculate the importance of all the frames:
\begin{equation}
  \textbf{V}_i^{\gamma} = {\rm softmax}(\textbf{V}_i),
\end{equation}
\begin{equation}
  \textbf{v}_{sum}^{\gamma j} = \sum_{i=1}^L \textbf{v}_i^{\gamma j}.
\end{equation}

We define the $\delta$ as the threshold. It is set as a constant. 
Then, we pick out the frame $j$, whose $\textbf{v}_{sum}^{\gamma j}$ is larger than $\delta$. 
% Then, we rank the $\textbf{v}_{sum}^{\gamma j}$ and pick out the top $\delta$. 
We view the selected frames as the text-consistent video part. The rest frames are the text-independent part.

\subsubsection{Multi-granularity Contrastive Learning}
Since there are no additional annotations, we apply the contrastive learning method for weakly supervised training of the cascaded action retrieval modules. 
For the Coarse-grained Action Retrieval module, we use the global level and clip level max-margin ranking loss to train it. Specifically, we view the randomly-selected video $\hat{\textbf{V}}$ and the paired sentence $\hat{\textbf{S}}$ as the negative samples. The proposal $\textbf{p}^1$ represents the feature of the whole video. We use it to calculate the ranking loss with the textual action feature $\textbf{s}_{act}$ to train the module from the global level, and the other candidate proposal features are used to calculate the clip level loss:
\begin{equation}
\begin{aligned}
  \mathcal{L}_{global} &= Max(0, \Delta_{global}-Sim(\textbf{p}^1, \textbf{s}_{act})+Sim(\textbf{p}^1, \hat{\textbf{s}}_{act})) \\
  &+Max(0, \Delta_{global}-Sim(\textbf{p}^1, \textbf{s}_{act})+Sim(\hat{\textbf{p}}^1, \textbf{s}_{act})),
\end{aligned}
\end{equation}
\begin{equation}
  \begin{aligned}
    &\mathcal{L}_{clip} = Max(0, \Delta_{clip}-\sum_{i=2}^K Sim(\textbf{p}^i, \textbf{s}_{act})+\sum_{i=2}^K Sim(\textbf{p}^i, \hat{\textbf{s}}_{act}\\&)) 
    +Max(0, \Delta_{clip}-\sum_{i=2}^K Sim(\textbf{p}^i, \textbf{s}_{act})+\sum_{i=2}^K Sim(\hat{\textbf{p}}^i, \textbf{s}_{act})),
  \end{aligned}
\end{equation}
where the $\Delta_{global}$ and $\Delta_{clip}$ are the margin values, the $\hat{\textbf{s}}_{act}$ is the feature of the negative text's action part. The $\hat{\textbf{p}}^i$ represents the clip level negative feature. Meanwhile, the $\hat{\textbf{p}}^1$ is the negative video feature. Then, we define the coarse-grained action retrieval module training loss as:
\begin{equation}
    \mathcal{L}_{coarse} = \mathcal{L}_{global} + \mathcal{L}_{clip}.
\end{equation}

For the Fine-grained Action Retrieval module, we need to train its text-consistent video frame selection ability. We view the proposal feature $\textbf{p}^c$ selected by the Coarse-grained Action Retrieval module as the negative sample. Then, the loss function is defined as:
\begin{equation}
  \begin{aligned}
    \mathcal{L}_{fine} = Max(0, \Delta_{fine}-Sim(\overline{\textbf{p}}^c, \textbf{s}_{act})+Sim(\textbf{p}^c, \textbf{s}_{act}))
  \end{aligned},
\end{equation}
where $\Delta_{fine}$ is a margin value and $\overline{\textbf{p}}^c$ is the output of the fine-grained action retrieval module.

\subsection{Hierarchical Multi-head Self Attention}
After distinguishing the action text description consistent and independent video frames, we analyse them using the \underline{M}ulti-head \underline{S}elf \underline{A}ttention (MSA). 
Traditional fully connected MSA lacks the spatio-temporal local sensitivity for dealing with the video. 

To address this problem, we devise the pyramid and shifted alignment mechanisms to fully use the spatial locality of the attribute semantics and the temporal locality of the action semantics. Based on these carefully designed structures, we propose the hierarchical MSA modules, containing Attribute-spatial MSA, Action-spatial MSA, and Action-temporal MSA. They fuse multi-modal information step by step from different semantics and spatio-temporal levels. This design avoids the interference between semantics, and 
simplifies the redundant attention, which interferes with model local sensitivity. 

Before presenting them, we first introduce some basic definitions. Assume that there are $I$ frames in one video and each frame has $J$ patches. We define the input video features as:
\begin{equation}
\textbf{F}_p=\{\textbf{f}_p^{1,0}, \textbf{f}_p^{1,1}, \textbf{f}_p^{1,2}, ..., \textbf{f}_p^{1,J}; \textbf{f}_p^{2,0}, ..., \textbf{f}_p^{2,J}; \textbf{f}_p^{I,0}, ..., \textbf{f}_p^{I,J}\}. 
\end{equation}
For convenience, we represent the patch features of the frame $i$ as: $\textbf{F}_p^i=\{\textbf{f}_p^{i,j}\}_{j=0}^J$. Meanwhile, $\textbf{f}_p^{i,0}$ is the frame (largest patch) feature.
In addition, with the application of the cascaded action retrieval module, the frames are divided into two parts, the text action-consistent frames $\textbf{F}_{pre}=\{\textbf{F}_{pre}^i\}_{i=1}^{I_{re}}$
and independent frames $\textbf{F}_{pir}=\{\textbf{F}_{pir}^i\}_{i=1}^{I_{ir}}$.

For the input language, the text features are represented as $\textbf{S}$. The action description part is $\textbf{S}_{act}$, and its information corresponds to the text action-consistent video frames. In addition, the rest sentence part (attribute part) consists of the subject and its non-action description, and is related to the whole video. We represent it as $\textbf{S}_{attr}$. 

\subsubsection{Attribute-spatial Layer}
For the attribute semantic content, we exchange the information within and between two modalities in the Attribute-spatial MSA at the spatial level. Specifically, we redefine the function $SeAttn(\textbf{Q}, \textbf{K}, \textbf{V})$ to get the pyramid attention functions:
\begin{equation}
  SeAttn_{pyK}(\textbf{Q}, \textbf{F}_p^i, \textbf{V}; W) = W(\textbf{Q}*(\textbf{F}_p^i)^T+\textbf{Q}*Avg(\textbf{F}_p^i)^T)\textbf{V},
\end{equation}
\begin{equation}
  SeAttn_{pyQ}(\textbf{F}_p^i, \textbf{K}, \textbf{V}; W) = W(\textbf{F}_p^i*\textbf{K}^T+Avg(\textbf{F}_p^i)*\textbf{K}^T)\textbf{V},
\end{equation}
where $W$ is a trainable parameter and $i$ is the index of the frame. $Avg$ makes average pooling for the $3*3$ area centered with each patch of the frame $i$. For edge patches, we only average the adjacent patches with the valid values, and ignore the padding patches. 
Then, introducing the pyramid attention functions into the function of the MSA, we get the new functions $MSA_{pyK}(\textbf{Q}, \textbf{F}_p^i, \textbf{V}; W)$, and $MSA_{pyQ}(\textbf{F}_p^i, \textbf{K}, \textbf{V}; W)$. 

For the language modal, we update the features of the attribute language part $\textbf{S}_{attr}$ with:
\begin{equation}
  \overline{\textbf{S}}_{attr} = MSA(\textbf{S}_{attr}, \textbf{S}_{attr}, \textbf{S}_{attr}) + MSA_{pyK}(\textbf{S}_{attr}, \textbf{F}_p^i, \textbf{F}_p^i).
\end{equation}

For the visual modal, we analyze the patch level features $\textbf{F}_p^i$ of the video frame $i$ with:
\begin{equation}
  \overline{\textbf{F}}_p^i = MSA_{pyQ}(\textbf{F}_p^i, \textbf{F}_p^i, \textbf{F}_p^i) + MSA_{pyQ}(\textbf{F}_p^i, \textbf{S}_{attr}, \textbf{S}_{attr}).
\end{equation}

\begin{table*}
  \caption{\label{sota_VidSTG}
  Compared with baselines on VidSTG. The $*$ represents the baselines using the MDETR as the pre-trained backbone, which is the same as the HERO. Overall \textcolor{red}{$1^{st}$} and \textcolor{blue}{$2^{nd}$} best in \textcolor{red}{red}/\textcolor{blue}{blue}. }
  \centering
  {\setlength{\tabcolsep}{1.5em}\begin{tabular}{c|cccc|cccc} \toprule[2pt]
  \multirow{2}{*}{\textbf{Method}} & \multicolumn{4}{c|}{\textbf{Declarative Sentence Grounding}} &\multicolumn{4}{c}{\textbf{Interrogative Sentence Grounding}}\\
          & 0.4   & 0.5   & 0.6   & Avg   & 0.4   & 0.5   & 0.6   & Avg  \\\midrule[1pt]
  \textbf{STPR} \cite{yamaguchi2017spatio}     & 41.8 & 30.1 & 19.5 & 30.4 & 33.0 & 27.4 & 18.3 & 26.2 \\
  \textbf{STGRN} \cite{zhang2020does}    & 46.1  & 35.0  & 23.6 & 34.9 & 39.6 & 30.6 & 22.1 & 30.8    \\
  \textbf{VOGnet} \cite{sadhu2020video}   & 45.3 & 36.5 & 27.5  & 36.4 & 41.4 & 33.2 & 23.6 & 32.7 \\
  \textbf{STVGBert} \cite{su2021stvgbert} & 58.1 & 49.6 & 40.9 & 49.5 & 48.5 & 39.5 & 30.8 & 39.6    \\\midrule[1pt]
  \textbf{VOGnet*}   & 49.3 & 42.3 & 33.7  & 41.7 & 44.5 & 37.9 & 31.3 & 37.9 \\
  \textbf{STVGBert*} & \textcolor{blue}{\textbf{66.9}} & \textcolor{blue}{\textbf{60.0}} & \textcolor{blue}{\textbf{50.3}} & \textcolor{blue}{\textbf{59.0}} & \textcolor{blue}{\textbf{52.8}} & \textcolor{blue}{\textbf{46.3}} & \textcolor{blue}{\textbf{40.0}} & \textcolor{blue}{\textbf{46.4}} \\ \midrule[1pt]
  
  \textbf{HERO} & \textcolor{red}{\textbf{70.9}} & \textcolor{red}{\textbf{64.5}} & \textcolor{red}{\textbf{56.0}} & \textcolor{red}{\textbf{63.8}} & \textcolor{red}{\textbf{56.0}} & \textcolor{red}{\textbf{50.6}} & \textcolor{red}{\textbf{44.9}} & \textcolor{red}{\textbf{50.5}}\\\bottomrule[2pt]
  \end{tabular}}
\end{table*}

\begin{table}
  \caption{\label{sota_VID-sentence}
        Compared with baselines on VID-sentence. The $*$ represents the MDETR is applied to these baselines as the pre-trained backbone.}
  \centering
  {\setlength{\tabcolsep}{1.3em}\begin{tabular}{c|cccc} \toprule[2pt]
  \textbf{Method} & 0.4 & 0.5 & 0.6 & Avg \\\midrule[1pt]
  \textbf{VOGnet*}   & 61.2 & 47.8 & 35.8 & 48.3 \\
  \textbf{STVGBert*}   & \textcolor{blue}{\textbf{63.6}} & \textcolor{blue}{\textbf{48.1}} & \textcolor{blue}{\textbf{40.3}} & \textcolor{blue}{\textbf{50.7}}\\\midrule[1pt]

  \textbf{HERO}   & \textcolor{red}{\textbf{67.5}} & \textcolor{red}{\textbf{53.3}} & 
  \textcolor{red}{\textbf{46.8}} & \textcolor{red}{\textbf{55.8}}\\\bottomrule[2pt]
  \end{tabular}}
\end{table}

\subsubsection{Action-spatial Layer}
For the action semantic content, the Action-spatial MSA is responsible for reasoning on them from the spatial level. Firstly, the attention function $SeAttn(\textbf{Q}, \textbf{K}, \textbf{V})$ is redefined as the shifted attention function:
\begin{equation}
  SeAttn_{shK}(\textbf{Q}, \textbf{F}_p^i, \textbf{V}; W) = W(\textbf{Q}(\textbf{F}_p^i)^T+\textbf{Q}(\textbf{F}_p^{i+1})^T+\textbf{Q}(\textbf{F}_p^{i-1})^T)\textbf{V},
\end{equation}
\begin{equation}
  SeAttn_{shQ}(\textbf{F}_p^i, \textbf{K}, \textbf{V}; W) = W(\textbf{F}_p^i\textbf{K}^T+\textbf{F}_p^{i+1}\textbf{K}^T+\textbf{F}_p^{i-1}\textbf{K}^T)\textbf{V}.
\end{equation}
Similar to the pyramid attention function, in this function, $W$ is learnable and $\textbf{F}_p^i$ is the patch level features of the frame $i$. We define the $\textbf{F}_{pre}^0=\textbf{F}_{pre}^{I_{re}}$ to guarantee the correct calculation of the frame $1$ and frame $I_{re}$. 
Then, the shifted MSA is obtained as $MSA_{shQ}(\textbf{Q}, \textbf{F}_p^i, \textbf{V})$ and $MSA_{shK}(\textbf{Q}, \textbf{F}_p^i, \textbf{V})$ based on the functions $SeAttn_{shQ}(\textbf{Q},  \textbf{F}_p^i, \textbf{V}; W)$, $SeAttn_{shK}(\textbf{Q}, \textbf{F}_p^i, \textbf{V}; W)$. 

The patch level features $\textbf{F}_{pre}^i$ of video frame $i$ are analyzed by:
\begin{equation}
  \overline{\textbf{F}}_{pre}^i = MSA_{skQ}(\textbf{F}_{pre}^i, \textbf{F}_{pre}^i, \textbf{F}_{pre}^i) + MSA_{skQ}(\textbf{F}_{pre}^i, \textbf{S}_{act}, \textbf{S}_{act}).
\end{equation}

What's more, the action text features $\textbf{S}_{act}$ are analyzed by:
\begin{equation}
  \textbf{S}_{act} = MSA(\textbf{S}_{act}, \textbf{S}_{act}, \textbf{S}_{act}) + MSA_{shK}(\textbf{S}_{act}, \textbf{F}_{pre}^i, \textbf{F}_{pre}^i).
\end{equation}

\subsubsection{Action-temporal Layer}
The Action-temporal MSA completes the information fusion for the frame-level features and the text features from the temporal perspective. We define the frame-level features as $\textbf{F}$. Meanwhile, the text action related video frames are $\textbf{F}_{ir}$, and the irrelevant video frames are $\textbf{F}_{re}$. 
Similar to the spatial level feature fusion modules, Attribute-spatial MSA and Action-spatial MSA, the Action-temporal MSA adopts different operations for these two types of video frames.

For the video features $\textbf{F}_{re}$, we update it by:
\begin{equation}
  \overline{\textbf{F}}_{re} = MSA(\textbf{F}_{re}, [\textbf{F}_{re}, \textbf{S}], [\textbf{F}_{re}, \textbf{S}]),
\end{equation}
where $[]$ represents the features concatenated.

For the video features $\textbf{F}_{ir}$, the reasoning function is:
\begin{equation}
  \overline{\textbf{F}}_{ir} = MSA(\textbf{F}_{ir}, [\textbf{F}_{ir}, \textbf{F}_{re}, \textbf{S}_{act}], [\textbf{F}_{ir}, \textbf{F}_{re}, \textbf{S}_{act}]).
\end{equation}

Notably, the information between $\textbf{F}_{re}$ and $\textbf{F}_{ir}$ is the single flow. It guarantees that the action text-consisted video frames help the detection of other video frames without being disturbed. 

\section{Experiments}

\subsection{Experiment Setup}
\subsubsection{Datasets}
Two video object grounding benchmarks, VidSTG and VID-sentence, are applied to evaluate the model performance:
(1) {\textit{VidSTG \cite{zhang2020does}}} This large-scale dataset is constructed based on the VidOR \cite{shang2019annotating} for video grounding benchmarking.
There are $10,000$ videos in VidSTG dataset and $99,943$ sentences with $55,135$ interrogatives and $44,808$ declaratives. $79$ types of objects appear in the videos described in these sentences.
In this paper, we use the official dataset split \cite{zhang2020does}.
(2) \textit{VID-sentence \cite{chen2019weakly}} This video object grounding dataset is a widely used benchmark. It is also based on VidOR dataset \cite{shang2019annotating}. The VID-sentence has $7,654$ video clips and $30$ categories. We follow the official split.

\subsubsection{Implementation Detail}
The model is implemented in Pytorch and trained using $8$ V100 GPU. We set model hyperparameter $\delta=0.8$. In most natural language video object grounding models, the pre-trained detection model is the parameter fundamental. In the same way as them, we choose the official pre-trained MDETR \cite{kamath2021mdetr} as the basis for the target detection of our HERO. We randomly resize the frames during video preprocessing, and set the maximum size to $640*640$. In addition, we also use the random size cropping and horizontal flip to deal with the input video. For the training process, we set the batch size as $1$, and the learning rate is $0.0001$, whose decay rate is $10$ for every $30$ epochs. Training epochs are limited to $90$. All experimental environments are deployed in Hikvision (\url{https://www.hikvision.com/en/}).

\subsubsection{Evaluation Metrics}
We adopt the same evaluation metrics as \cite{chen2019weakly}. In detail, we compute the Intersection over Union (IoU) metric using the ground-truth and predicted bounding box of each frame. The threshold $\alpha$ is set. If the average IoU for the whole video exceeds $\alpha$, the prediction is viewed as "accurate". We set the $\alpha$ as $0.4$, $0.5$, and $0.5$ during testing.

\subsubsection{Baselines}
Several state-of-the-art models are adopted as the baselines to compare. In detail, (1) fully supervised video grounding models: \textbf{STGRN} \cite{zhang2020does}, \textbf{STVGBert} \cite{su2021stvgbert}, and \textbf{VOGnet} \cite{sadhu2020video}; (2) other widely known method: video person grounding \textbf{STPR} \cite{yamaguchi2017spatio}.

\subsection{Performance Comparision}
In Table~\ref{sota_VidSTG} and Table~\ref{sota_VID-sentence}, we present the video object grounding experiment results on VidSTG and VID-sentence datasets, respectively.
The following conclusions can be drawn from the tables:

\begin{itemize}

  \item Our HERO model performs best on two benchmark datasets compared with previous methods. Remarkably, in comparison to the previous state-of-the-art, STVGBert, the HERO model significantly improves performance (Avg) from nearly $59.0\%$/$46.4\%$/$50.7\%$ to $63.8\%$/$50.5\%$/$55.8\%$ on VidSTG and VID-sentence datasets, which proves the effectiveness of HERO on the video object grounding.

  \item The video person grounding method STPR, that belongs to the other domain and is extended to the video object grounding setting, performs worse on the video object grounding benchmark. It is because the method lack domain-specific knowledge and could not effectively process the spatio-temporal cross-modal information under the setting of this task.

  %%异常
  \item In the original paper, the baselines are implemented with different backbone networks, compared with ours. To further prove the improvement of HERO coming from the reasonable model design, we re-implement two performing best baselines (VOGnet and STVGBert) using the same object detection backbone, MDETR, as HERO.
  Our HERO model still outperforms the best performing baseline STVGBert by over $4\%$ points (avg) on all two datasets, regardless of the new backbone's improved performance.
  It further demonstrates the superiority of our proposed model in improving the alignment effect of the cross-modal. The improvement can be attributed to the end-to-end modeling. This allows different modules to benefit from one another simultaneously. In addition, the Coarse-to-Fine Action Retrieval module
  aligns the fine-grained video semantics (often with multiple actions) and the coarse-grained text semantics (often with single action).
  Our proposed new multi-head self attention layers (Attribute-spatial MSA, Action-spatial MSA, and Action-temporal MSA) effectively exploit the spatial locality of attributes and the temporal locality of actions.

\end{itemize}

{\setlength{\tabcolsep}{0.7em}\begin{table}[t]
  \caption{\label{ablationstudyVIDS}
Ablation study of the effect of each module in HERO on the declarative sentence grounding of the VidSTG dataset.}
  \centering
  \begin{tabular}{cccc|cccc}\toprule[2pt]
  $\mathbf{\theta_{hier}}$ & $\mathbf{\theta_{retr}}$ & $\mathbf{\theta_{pyr}}$ & $\mathbf{\theta_{shi}}$ & 0.4   & 0.5   & 0.6   & Avg  \\\midrule[1pt]
  &  &  &  & 68.3 & 61.4 & 51.6 & 60.4 \\
  \checkmark&  &  &  & 69.2 & 61.9 & 52.2 & 61.1    \\
  \checkmark&  & \checkmark &  & 69.5 & 62.0 & 53.0 & 61.5 \\
  \checkmark&  &  & \checkmark & 69.7 & 62.4 & 54.1 & 62.0 \\
  \checkmark& \checkmark &  &  & 70.0 & 63.4 & 54.9 & 62.8 \\
  \checkmark& \checkmark & \checkmark &  & \textcolor{blue}{\textbf{70.7}} & 63.3 & 55.3 & 63.1 \\
  \checkmark& \checkmark &  & \checkmark & 70.2 & \textcolor{blue}{\textbf{64.0}} & \textcolor{blue}{\textbf{55.7}} & \textcolor{blue}{\textbf{63.3}} \\
  \Checkmark& \Checkmark & \Checkmark & \Checkmark & 
  \textcolor{red}{\textbf{70.9}} & 
  \textcolor{red}{\textbf{64.5}} & 
  \textcolor{red}{\textbf{56.0}} & \textcolor{red}{\textbf{63.8}}\\\bottomrule[2pt]
  \end{tabular}
  \end{table}
}

\subsection{Ablation Study}
In order to fully evaluate the effectiveness of the HERO model, we would like to know how different building blocks contribute. We accomplish this by removing several components from HERO and constructing the new architectures.
The investigated modules include the coarse-to-fine action retrieval module, three hierarchical multi-head attention modules (Attribute-spatial MSA, Action-spatial MSA, and Action-temporal MSA), pyramid alignment module, and shifted alignment mechanism. For convenience, we use $\theta_{retr}$, $\theta_{hier}$, $\theta_{pyr}$, and $\theta_{shi}$ to represent them.
Results on VidSTG are shown in Table~\ref{ablationstudyVIDS}, respectively. Notably, the other modules cannot work independently without the hierarchical fusion modules. Thus, we don't conduct ablation studies for them alone.  
There are several observations:

\begin{itemize}

\item After adding the Attribute-spatial MSA, Action-spatial MSA, and Action-temporal MSA, the model performs better. It reveals that three MSA modules effectively exclude the interference of redundant attention by extracting the spatio-temporal correlation separately, compared with the fully connected multi-head attention layer. 

\item The performance is better to use multiple components of the HERO architecture together rather than a single component architecture. The experiment results demonstrate that our proposed model components complement each other for the video object grounding. When we apply them together, the model acquires stronger feature relationship extraction and analysis capabilities

\item The model performs worse when any component of the model is removed
It demonstrates the effectiveness and necessity of our coarse-to-fine action retrieval module, three hierarchical fusion MSA modules, pyramid alignment mechanism, and shifted alignment mechanism. 

\end{itemize}

\begin{figure}[t]
  \centering
  \includegraphics[width=\linewidth]{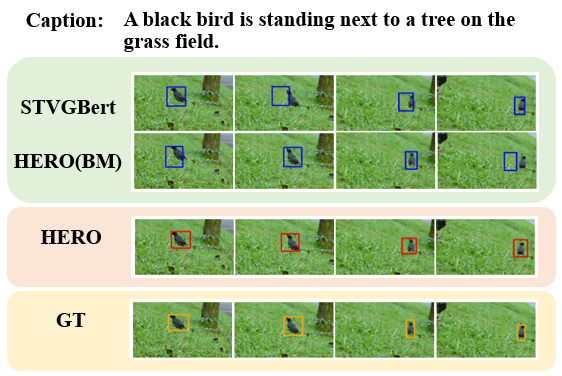}
  \caption{Examples of video object grounding result visualization. The HERO(BM) represents the base model defined in the ablation study. The GT is the target result.} 
  \label{figure4-1}
\end{figure}

\subsection{Case Study}
In order to prove the ability of the HERO in visuals, we conduct a case study. In particular, we randomly sample a video from the datasets.

To fully demonstrate model performance on the sample,
we compare the HERO model with the best performance baseline method, STVGBert, and the fundamental ablation model of the HERO, whose the Coarse-to-Fine Action Retrieval module and Hierarchical MSA module of the model are removed. According to the Figure~\ref{figure4-1}, there are the following key findings:

\begin{itemize}
  
  \item By adding the consisted modules of the HERO, the model predicts more accurately. It demonstrates that these components of our model improve the degree of cross-modal alignment so that the model has better detection ability.
  
  \item Compared with the performance best baseline method, STVGBert, our HERO detects the most accurate one from all the candidate objects and predicts more accurate bounding boxes. It reveals that our model has better representation analysis ability.

\end{itemize}

\section{Conclusion}

In this paper, we analyse two main challenges of the multi-modal \underline{V}ideo \underline{o}bject \underline{G}rounding (\emph{VoG}) task: (1) Incomplete Cross-modal Semantic Correspondence. (2) Insensitive Spatio-Temporal Locality.
To tackle these problems, we propose an end-to-end \underline{\textbf{H}}i\underline{\textbf{E}}rarchical sptio-tempo\underline{\textbf{R}}al reas\underline{\textbf{O}}ning (HERO) model with action correspondence for the VoG task. The HERO firstly retrieves the textual action-consistent semantics from the input video with weakly supervised contrastive learning from multiple granularities. Then, it reasons the divided multi-modal information step by step according to the different semantics using carefully designed locality-sensitive multi-head self attention layers, thereby fully utilizing the attribute (\ie, object) spatial locality and the action temporal locality.
The experiment results on two benchmark datasets demonstrate the significant effect of our HERO model.

\section{Acknowledgments}
\begin{sloppypar}
Our research is funded in part by the Program of Zhejiang Province Science and Technology (No.2022C01044), NSFC (No.62037001), Key R\&D Projects of the Ministry of Science and Technology (No.2020YFC0832500), Zhejiang Natural Science Foundation (No.LR19F020006), and National Key R\&D Program of China under Grant (No.2020YFC0832505). In addition, our research is funded in part by Hangzhou Hikvision Digital Technology.
\end{sloppypar}

\bibliographystyle{ACM-Reference-Format}
\bibliography{sample-base}

\end{document}